\documentclass[aps,showpacs,preprint]{revtex4}
\usepackage{graphicx}% Include figure files
\usepackage{dcolumn}% Align table columns on decimal point
\usepackage{bm}% bold math

\begin{document}

\title{Scanning Thermal Microscope Study of a Metal Film Under Current Stressing: Role of Temperature Inhomogeneity in Damage Process \\} 

\author{Achyut Bora\footnote[1]{Electronic mail: achyut@physics.iisc.ernet.in}$^1$ and A. K. Raychaudhuri\footnote[2]{Electronic mail: arup@bose.res.in}$^{1,2}$}
\address{$^1$Department of Physics, Indian Institute of Science,  Bangalore 560 012,  India\\
$^2$DST Unit for Nanoscience, S.N.Bose National Centre for Basic Sciences, Salt Lake, Kolkata 700098, India}

\date{\today}

\begin{abstract}
We report  direct observation of the evolution of local temperature inhomogeneity and the resulting atomic migration in a metal film (Ag on Si) stressed by a  current by using a Scanning Thermal Microscope that allows simultaneous temperature mapping and topography imaging. The experimental observation is analyzed using a model based simulation. The experimental observation and the simulation show that due to current stressing the temperature of the film becomes significantly inhomogeneous over time (with local temperature deviating strongly from the mean). This creates local stress as well as local temperature gradient that lead to mass migration in addition to the electromigration. We show that the local temperature  inhomogeneity serves as one of the main agents for local atomic migration  which leads to change in film microstructure. The migration leads to damage and eventual failure as simultaneously monitored by \textit{in-situ} resistance measurement. 
\end{abstract}

%\keywords{ scanning thermal microscopy,current  stressing, electromigration, thermomigration}
\maketitle

\section{Introduction}
In presence of large electric current through a metal film, ions in the film are acted upon by a number of  forces (in addition to the direct electrostatic force) that lead to migration of ions . These forces can be due to momentum transfer from electrons to the ions leading to electromigration (EM)  or due to stress (stress migration) or  stress resulting from thermal gradient~\cite{TM1,huntington1,Blech2,emrev1,emrev2,StressTM1}. Whatsoever may be the cause of this mass  migration, it leads to change in film microstructure and formation of voids or hillocks  leading to complete damage when the current density is very high. In this paper we report an important experimental result which clearly shows that the initiation of mass migration in a current carrying metallic film is associated with  development of local  temperature  inhomogeneity (spatial) which acts as the sites for nucleation of atomic migration and further microstructural evolution depends substantially on the evolution of spatial temperature inhomogeneity during the entire damage process. This shows that in addition of the average temperature of the film, the local temperature (which is different from the average) plays an important role. We did this by imaging the temperature variation of the surface of a current carrying metal film using a scanning thermal microscope (SThM) which allows us also to generate a simultaneous topography map in combination of \textit{in-situ} measurement of the resistance of the film. It has already been demonstrated that SThM can be used as a powerful thermal imaging tool to map the temperature distribution due to joule heating effect in metal lines carrying high current~\cite{Majumdar}. The dependence of atomic migration through metal films on the equilibrium temperature of the film and the joule heating effect causing change in average film temperature  have been investigated before~\cite{Colm,Joule1}. However, to understand the dependence of the damage process on thermal distribution, it is very important to study the correlation between evolution of microstructure and the spatial inhomogeneity of temperature as a function of time. In particular, development of local temperature gradient ($\nabla T$) and its impact on the mass migration are important factors and it will be important to follow the evolution of these factors as the damage process evolves with time. To our knowledge, this important issue has not been addressed to in any earlier study. Our experiment, being based on simultaneous topographic imaging  and spatially resolved temperature mapping along with \textit{in-situ} resistance measurement with the progress of the damage process, directly addresses these issues. Imaging techniques like Infrared Microscopy~\cite{kondo}, Laser-reflection thermometry~\cite{ju}, SThM~\cite{Majumdar} and Scanning Joule expansion microscopy~\cite{igeta} have been used in the past to image the thermal inhomogeneity of metal interconnects with different degrees of spatial resolution. However, the evolution of the inhomogeneity, local temperature gradient and its direct correlation to the mass migration as the damage process evolves (established by \textit{in-situ} monitoring of the resistance)  have not been done before.  In addition to the direct experimental observation we also carry out a simulation based on a model to analyze the effect  of the local temperature inhomogeneity on the mass migration process and thus gain understanding of the observed results. 

\section{Experimental techniques}
Experiments were carried out in Ag films (thickness $\approx 0.160 \mu m$)  as test samples (dimension $\approx 20 \mu m$ square). The films were  prepared by thermal evaporation on Si (held at $175^\circ C$) at a base pressure of $10^{-8} mbar$ and post-growth annealing at $250^\circ C$ for 6 hours in the same vacuum. The XRD of the films show that the films were mostly oriented in $(111)$ direction and well textured. The room temperature resistivity ($\rho_{300K}$) and residual resistivity ratio ($\rho_{300K}/\rho_{4.2K}$) for the films were typically around $1.65 \mu \Omega cm$ and $\sim8$ respectively. 

FIG~\ref{SThMSchematic} shows the schematic for \textit{in-situ} resistance and SThM as well as topography measurements. The SThM and topography measurement was based on a contact mode Atomic Force Microscopy (AFM) technique with an additional temperature measurement unit (SThM module) and using the cantilever of AFM itself as a thermometer. We used a commercial AFM from Veeco Metrology Group ( Model: Autoprobe CP Research) and a cantilever  with a built-in resistive thermal element made of Pt/10\% Rh alloy which acts as a probe for scanning thermal microscope~\cite{Veeco}. A mirror, cemented on the cantilever, acted as the reflector for the laser light to measure the deflection of the cantilever. To measure the temperature of the sample at the point where the tip of the probe touches it, the probe was connected as an arm of a balanced Wheatstone bridge powered  by a constant current from SThM module.    
The error voltage, $V_{out}$ (see FIG~\ref{SThMSchematic}), from the bridge varies linearly with local temperature (T) of the  point where the tip touches the film and is given as
\begin{equation}
V_{out}=V_0+i\alpha R_0 C(T-T_0)
\label{SThMWheastoneEq1}
\end{equation}
where,  
$V_0$ = bridge voltage at the reference temperature, 
$i$ = current applied to the probe,
$\alpha$ = temperature coefficient of resistance of the probe material,
$R_0$ = resistance of the probe at a reference temperature $T_0$ and
$C$ = calibration gain of the SThM module.
%~\ref{SThMWheastoneEq1}. 
Before carrying out measurement on the actual sample, the SThM probe was calibrated against $V_{out}$. This was done by  using a Pt100 film thermometer as a sample which was placed on a heater and recording $V_{out}$ for  different reference temperatures as measured by the  Pt100 film thermometer which is serving as the sample. From a  fit through these calibration points the absolute temperature could be calculated according to the equation,
\begin{equation}
T=(T_0-\frac{1}{m}V_0)+\frac{1}{m}V_{out}
\label{SThMWheastoneEq2}
\end{equation}
%~\ref{SThMWheastoneEq2}
for any value of $V_{out}$. From the calibration curve we find the slope  $m=8.3 mV/K$.  
The spatial resolution of the thermal (local T) images was limited by the finite size of the  tip to $\approx 200nm$.  The test sample size was kept in the range of $\approx 20 \mu m$ due to this spatial resolution . We applied  a direct stressing current (dc) of density $J_{dc}=2\times 10^{7} A/cm^{2}$ across the current leads and the voltage drop was recorded to measure the resistance. The damage process was monitored along with imaging by directly recording the resistance (R)  at regular time interval automatically until the film got damaged.  With this experimental setup we could investigate systematically the evolution of the damage  process with \textit{in-situ} AFM and SThM.

\section{Result}
In the figure~\ref{SThMEMplot} we plot the evolution of the resistance of a film as a function of time with the progress of the damage process. Resistance is plotted in logarithm scale to enhance the initial lower resistance regions.  It was observed that initially the resistance increased very slowly for first $105~hours$ after which (marked as $t_1$ in the plot) it showed rapid increase. Before reaching the point $t_1$, incubation of  resistance followed $R=R_0e^{p_1t}$ kind of trend, where, $R_0$ is the initial resistance of the film and $p_1\approx 2\times 10^{-3} h^{-1}$ is the slope of the $logR-t$ plot. After this point also increase in resistance followed a similar trend but with a  much larger slope $p_2\approx 3.6\times 10^{-2}h^{-1}$, showing a rapid increase. This trend was seen to be maintained  till around $120~hours$  (marked as $t_2$ in the plot) after which also resistance kept increasing at a similar average rate. In the late stage of damage process, as shown in the FIG~\ref{SThMEMplot}, there are steps in resistance changes. We generally take the SThM and topography images in these regions because the resistance stays constant over a time period. The steps are marked as $t_0$, $t_1$.....$t_7$ in the figure.

In the FIG~\ref{TopoSThM}, three representative topography, SThM images and temperature profiles are shown.  The data obtained at different stages of current stressing process, are shown in three different rows. The progress of the damage process can also be seen from the change in R of the film with the time of current stressing (FIG~\ref{SThMEMplot}). Resistance change in the FIG~\ref{SThMEMplot} and the AFM based imaging were taken simultaneously. The first figure in the row (a) shows the topography of the whole film including contact pad, while in all the other images only the part of film, marked by a square, is shown to elaborate the changes that occured as a result of current stressing. Row~(a) corresponds to the film at unstressed condition (at $t_0$). The line-scan is a typical example of temperature inhomogeneity in an unstressed film and shows the noise limited resolution. As the damage process progresses one sees there appearance of relatively hot areas surrounded by relatively cold contours as indicated by arrows in the SThM image in row~(b), which was taken after 146 hours (at $t_3$) of current stressing. A typical line-scan on SThM image shows that the spatial temperature fluctuation has gone up. Simultaneous contact mode AFM mapping shows that these local hot spots act as the regions where the films's microstructure has changed considerably. This becomes more visible (see row~(c)) in later stage of the damage process where  there were sufficient changes in microstructure and the film becomes rough (with formation of  hillocks and voids) at these positions and a more non-uniform temperature profile develops. The row~(c) images were taken after 163 hours (at $t_5$) of current stressing. As the damage process progresses, the temperature over the film become more non-uniform  and a larger part of the film gets damaged. [\textit{Note: The line scans, shown in the FIG~\ref{TopoSThM} at different stages of current stressing, were taken at the same sites to show gradual evolution of temperature inhomogeneity.} 

The result of \textit{in-situ} SThM measurement qualitatively shows that the progress of damage process is accompanied by the increase in surface roughness, enhanced local temperature variation and also an enhanced average temperature. A quantitative evaluation of these observations can be obtained from the development of these parameters with the progress of the damage process. Correlations among these parameters will lead to a more clear picture. To achieve this, we calculate rms roughness of the film ($\sigma (h)$), the average  temperature ($<T>$) and the relative variances in temperature ($<(\Delta T)^2>/<T^2>$) for an arbitrarily chosen part of the film at different stage of the damage process. The averages are taken over the whole surface of the film. We have marked such an area with a square in the topography image in the row~(a)of the FIG~\ref{TopoSThM}. The dimension of this area is $12\mu m \times12 \mu m$. All the further calculations presented here were done for this area. 

To calculate the rms roughness of the film we find out the hight profiles along some lines over the selected part of the film. The standard deviation of the height profile gives the rms roughness along a line. We draw 60 numbers of such line in horizontal direction and 60 lines in the vertical direction. Thus each line is separated by $\approx 200 nm$, a value close to the resolution of the SThM tip. For each stage of damage  we calculated the rms roughness averaged for all the lines. This rms roughness ($\sigma (h)$) (averaged for the area) for the stages marked as $t_0$, $t_1$, $t_2$....$t_7$ in the FIG~\ref{SThMEMplot} are plotted in the FIG~\ref{ResTimeRMSmeanT}(a) as a function of time along with the resistances of the film at the respective stages. From this plot increase in $\sigma (h)$ with the time of current stressing can be clearly seen. Since the roughness of surface has  resulted from the  mass flux in the film due to the migration process, the quantities $\sigma (h)$ at any stage is a measure of the mass migration of the film that  occured at the corresponding stage due to the current stressing. Initially $\sigma (h)$ increased very slowly, but became very rapid when the resistance change is also rapid. The simulation, described later on, shows that this happens when there is  significant mass migration. (In the example of the film shown here this happens after 146 hours of stressing). Towards the end of the experiments the $\sigma (h)$ increased by more than one order of magnitude.

To study the evolution of average temperature with the progress of the damage process, we calculate mean temperature ($<T>$) for the same part of film in the same way as was done for calculation of $\sigma (h)$. Here we take the thermal map of this part of film and find out the temperature profile along 60 numbers of horizontal lines and 60 numbers of vertical lines. Some typical temperature line profile are already shown in the FIG~\ref{TopoSThM}. The mean temperatures ($<T>$) for various stages of damage, calculated from the mean temperature for each line and averaged over the all lines give average temperature of this film area at corresponding stages of damage. The evolution of $<T>$ is plotted in the FIG~\ref{ResTimeRMSmeanT}(b) as a function of stressing time. With the advances of the stressing time $<T>$ increases, initially little slowly but rapidly after 120 hours of stressing. Relative variances of temperature ($<(\Delta T)^2>/<T^2>$) for any stage was obtained from calculating the variance of temperature for each line and averaging over all the lines and dividing the resultant quantity by the average of the squares of temperature along all the lines. The quantity $<(\Delta T)^2>/<T^2>$ represents spatial inhomogeneity of temperature over the selected area irrespective of the background temperature. Change in relative variances in temperature is shown in the FIG~\ref{ResRelVarRMSgrad}(a) as function of time. Corresponding resistance values are also shown there. Almost, throughout the stressing time, all these quantities were seen to increase with advances of the damage. 

\section{Discussion and Simulation}

\subsection{General discussions}
The results of the experiment  show that local temperature variation that is created on current stressing is an important physical parameter in the current stressed film and it plays an important role in th eventual failure of the film.The increase in average temperature of the film is due to the joule heating effect. Due to microstructural change in the film caused by atomic migration during the damage process, the local resistance changes and this leads to spatial variation of the  current density. This will lead to local variation in heat dissipation  which will give rise to the observed spatial variation of local temperature in the film. This particular issue that local resistance inhomogeneity leads to inhomogeneity in local heat dissipation and hence the local temperature variation is also borne out by the model simulation discussed below. With the advances of the stressing time, increased surface roughness of the film with introduction and development of voids or hillocks in the film, such temperature distribution is expected to show stronger spatial variation, irrespective of the background average temperature.This is what was observed in this experiment. Comparing evolution of $<\Delta T^2>/<T^2>$ with that of $\sigma (h)$, it could be seen that for most of the stressing time, relative spatial variation of temperature increased along with the increased surface roughness which became very rapid towards the end of the damage process with increase in value by more than two orders of magnitude. After the damage commences the local mass migration that results from the local temperature variation leads to enhanced average temperature (because larger heat dissipation resulting from larger $R$) as well as enhanced local gradient leading to more mass migration. Eventually there is a ``run away" situation in all the parameters and the film is completely damaged.

In a current stressed film, temperature has a great influence on the mass migration and the local atomic flux. Particularly, atomic migration resulting from local temperature gradient or thermomigration becomes one of main components of the net mass migration, while the other components are electromigration and stress- migration. The component of mass flow arising from thermal gradient scales with $\frac{1}{T^2}\nabla Te^{\frac{-E_a}{k_{\beta}T}} $ where $E_a$ is the activation energy for diffusion~\cite{TM2}. The temperature map obtained from the SThM images at different stages of the damage process allow us to calculate the local temperature gradient. We obtained the magnitude of the temperature gradient ($\mid\nabla T\mid$) by averaging over the same regions that we used to calculate rms roughness. To calculate $\mid\nabla T\mid$ we used same temperature profiles for all the lines used as described above. We obtained $\frac{\partial T}{\partial x}$ and $\frac{\partial T}{\partial y}$ from the horizontal lines and vertical lines respectively for each of their cross-point and then we calculated the local value of $\mid\nabla T\mid$ at that point by taking square root of $(\frac{\partial T}{\partial x})^2+(\frac{\partial T}{\partial y})^2$. When this quantity was averaged over all cross-points, $\mid\nabla T\mid$ gave the magnitude of the temperature gradient averaged for the selected region of the film. In the FIG~\ref{ResRelVarRMSgrad}(b) we have plotted $\mid\nabla T\mid$ as function of time. To show development of surface non-uniformity with the evolution of temperature gradient, we have plotted $\sigma (h)$ also in the same figure. It can be clearly seen from this plot that the local gradient, $\mid\nabla T\mid$  increases monotonically  with time and that the rms roughness follows the same trend. Thus, the local temperature distribution as well as the local gradient are related and both have significant effect on the migration process. This particular observation that local temperature gradient controls in a significant way the direction of local direction of mass migration is an important out come of our experiment. The model described below provides support to this explanation.

The effect of local temperature distribution on the mass flow can be mediated via some other related underlying physical processes. Particularly local variation in temperature produces a mechanical stress field in the film due to mismatch of thermal coefficients of expansion of the film material (Ag) and the substrate (Si). The stress field generated thus also acts as an additional strong driving force for atomic migration. Though it is not possible to single out the measure of damage resulting from the temperature gradient field alone, the results of \textit{in-situ} SThM have shown clearly that local temperature variation has strong influence on the local atomic mass flux.

\subsection{Model of the process and simulation}

The damage process of current stressed metal film is highly complicated process where various phenomena dependent of number of parameters like current density, operating temperature, thermo-mechanical properties of the film and other materials in physically contact with it. It is also important to note that the film cannot be considered as smooth. It has topological features and also local variation in resistance. It is thus not possible to obtain an accurate analytical result on the evolution of damage process . However, with a good model  the damage process can be simulated. As a part of our present investigation, we used a realistic model and  carried out a simulation in order to gain an understanding of the microstructural evolution as well as the local  temperature evolution of the film as the  resistance changes as a result of the current stressing. We will see that the model, though simple captures most of the features observed in the experiment. Below we describe the model briefly and present the results. The detailed model and the computational methods are beyond the scope of this paper and have been presented separately~\cite{achyutThesis}.

 In this simulation we used the initial topography  of the film as revealed by the AFM image as the  input.  This was done by taking AFM topography image of pristine film. Then the film was modeled as a resistance network (see the FIG~\ref{FilmFEMResNetwork}) and the film/substrate system was modeled as finite element (FIG~\ref{FilmFEM}). The  value of the each local  resistance element was obtained from the  height profile of the AFM image. We made the assumption that on the average the resistivity is homogeneous across the film but the local resistance variation arises from the local height variation (that then controls the cross-section of the local resistance element) which is captured by the actual AFM image. The equivalent resistance  of the film was obtained by solving the network. This equivalent resistance was then compared with the experimentally measured resistance of the film and a scaling factor was obtained. The scaling factor was then used to scale the values of  all the resistance elements of the network. In this way we got a more realistic value of the local resistance elements that comprise the network and that is compatible with the resistance of the actual film. 

When the network is biased with a constant current source, the current through the each resistance element was calculated. Each resistance element also behaved as a source of local heat input due to joule heating. The finite element modeling of the film was done to calculate correctly the local temperature at any point of the film. The height of a finite element for the film at any location was taken according to the film's height at that location whereas the height of all the elements for the substrate were kept fixed throughout the simulation. Using finite element analysis (FEA, described briefly in the appendix) the local temperatures for each node of all elements were calculated for the film and the junction between the film and the substrate. The corresponding stress due to mismatch of thermal expansion coefficients (TEC) of the film and the substrate were calculated. Using the just  obtained local temperature and stress, net mass flux arising from combined contribution of  `electromigration', `stress-migration' and `thermo-migration' was calculated. For this we used the formula:
\begin{equation}
J_a=\frac{D_\circ C}{k_\beta T}(Z_a^\ast e\rho J-\Omega\nabla\sigma -\frac{Q^\ast}{N_AT}\nabla T)exp[-E_a/k_\beta T]
\label{atomicFluxEq}
\end{equation}
where, $J_a$ is the atomic flux, $D_\circ$ is the prefactor of diffusivity, $C$ is atomic concentration, $Z_a^\ast$ is effective valence of the material, $\rho$ is resistivity of the film, $J$ is current density, $\Omega$ is atomic volume, $\sigma$ is the stress, $Q^\ast$ is heat of transport, $N_A$ is the Avogadro's number and rest of the symbles have usual meaning. For Ag film, we took values of $D_\circ =6.4\times 10^{-7}~m^2/s$~\cite{brune95} for self surface-diffusion, $Z_a^\ast = -21$~\cite{huntington1}, $\rho=1.65\times 10^{-8}~\Omega .m$ (experimentally obtained), $\Omega =  1.7\times 10^{-29}~m^3$, $Q^\ast=0.5~eV$ and $N_A=6.022\times 10^{23}$. We used $J = 2\times 10^7~A/cm^2$ and kept $E_a$ as adjustable parameter.

The resulting mass migration was then used to recalculate the new  height profile of the film. The new local height profile becomes the input for the next simulation steps. At the end of each simulation step we can obtain the new resistance of the film, the average temperature, the local height variation (which we expressed as the rms roughness) and the local temperature variation that we expressed as the rms temperature fluctuation averaged over the film surface. These are also the experimentally determined observable. This process was followed in every simulation step. Thus, in our simulation we could see evolution of microstructure and local variation in temperature along with evolution of the film resistance  as a function of time (which is the number of simulation steps)as the damage process  progress as a result of the current stressing.

The results of the simulation (plotted as number of simulation steps) are shown in the FIG~\ref{SimResTimeRMSmeanT} and FIG~\ref{SimResTimeRelRMSgrad}. The results of simulation can be compared with the experimental results shown in  FIG~\ref{ResTimeRMSmeanT} and FIG~\ref{ResRelVarRMSgrad}. In the FIG~\ref{SimResTimeRMSmeanT}(a) and~\ref{SimResTimeRMSmeanT}(b) we show the evolution of rms roughness ($\sigma (h)$) and the average temperature ($<T>$) alongwith the evolution of resistance as obtained from the simulation. In the FIG~\ref{SimResTimeRelRMSgrad}(a) we show the evolution of resistance of the film along with the change in relative variances of temperature. The evolution of surface roughness  and  that of the temperature gradient can be seen from the FIG~\ref{SimResTimeRelRMSgrad}(b). The simulation showed that the surface roughness, mean temperature, temperature gradient and the resistance increase slowly with time initially , but at very rapid rate after some instance when the resistance starts to change at a rapid rate. This result  is in conformity with what has been observed experimentally.The simulation not only reproduces the experimental observation qualitatively it also has quantitative validity. As an indicator we can compare the numbers when as result of the migration the resistance $R$ of the film reaches $1 \Omega$. In experimental observation the parameter surface roughness $\sigma (h)$
reaches $\approx 0.017~\mu m$. The simulation shows a value of $\approx 0.023~\mu m$. The average temperature $<T>$ at that time is $\approx 339~K$. The corresponding number from the simulation is $\approx 321~K$. The average local temperature gradient $\mid\nabla T\mid$ at that instance obtained from the experiment and from the simulation are $0.95~K\mu m^{-1}$ and $1.13~K\mu m^{-1}$ respectively. The Relative variances of temperature ($<(\Delta T)^2>/<T^2>$) for that  stage from the experiment is $2\times 10^{-6}$ and from the simulation it is $4\times 10^{-6}$. The two local thermal parameters obtained from simulation  though similar to the experimental result are however not the same. This reflects the complexity in modeling the underlying local heat transport process. Nevertheless we find that the simulation reflects in a significant way the process of current stressing.This is encouraging because there are no adjustable parameters in the simulation except the activation energy $E_a$. 

The simulation establishes the basic understanding that due to increase in the local temperature gradient, enhanced mass migration was observed resulting in an  increase of  surface roughness also. The simulation shows that at this instance there are   substantial mass migration in the film. The simulation shows that this increases the resistance values of the local elements leading to an enhancement of the average temperature are controlled largely by the  local temperature gradient as well as the local temperature inhomogeneity as expressed in the local rms temperature fluctuations (in addition to the average temperature T).  

The simulation shows that the temperature enters in two ways. First, since the migration is an activated process, enhanced average temperature accelerates the process. Second, an enhanced local gradient of $T$ also enhances the atomic flux.

\section{Conclusion}
To summerize, in this investigation, we have carried out a systematic study on time evolution of local temperature variation as well as the local topology variation  of  a thin metal film subjected to high current stressing using AFM and SThM. We also measured in-situ the resistance change to monitor the progress of damage of the film under current stressing.   We found that although the film has uniform temperature initially, at some point of time there appear regions with relatively large temperature inhomogeneity with hotter zones surrounded by colder contours. The migration process nucleates in these regions and significant atomic migration results here as the time progresses. As the time progresses, new hotter and colder zones with higher local temperature difference appear. Our experiment demonstrates a close correlation among the atomic migration, the variation of local temperature and microstructural variation like surface roughness. The experimental observation was analyzed using a model based simulation . The results of the simulation show that  the local heating controls the stress as well as the thermal gradient ($\mid\nabla T\mid$). The migration due to local stress and the thermal migration  add upto the electromigration induced mass migration and  create the total the mass migration. This leads to damage of the film. Our investigation   shows the predominant role of not only the average temperature growth but also of the local spatial temperature variations  that determines the nature of net mass flow into or out of a region of the film under current stressing leading to its damage and eventual failure.  

\begin{acknowledgments}
A.B.  thanks University Grants Commission, India for a fellowship and A.K.R thanks Department of Science and Technology and Department of Information Technology  Govt. of India for sponsored projects.
\end{acknowledgments}

\appendix*
\section{A}
\begin{figure}
\vspace{6 cm}
\includegraphics[width=8cm]{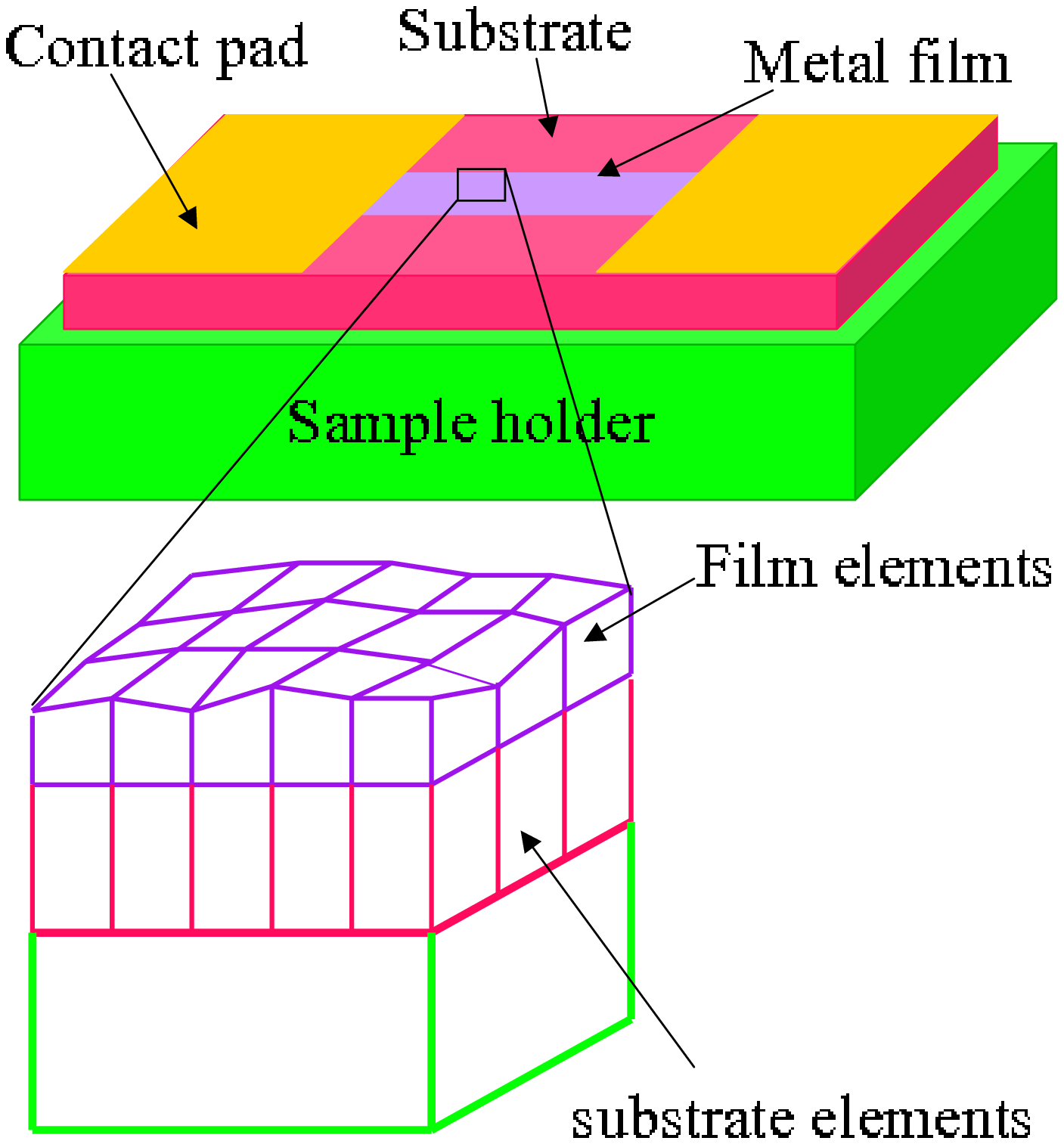}
\caption{}
\label{FilmFEM}
\vspace{6 cm}
\end{figure}

Finite element analysis was aimed to calculate temperature distribution throughout the film and even in the substrate at every simulation step by solving the heatflow equation 
\begin{equation}
\rho_m s\frac{\partial T}{\partial t} = \nabla .(k\nabla T)+Q
\label{heatFlowEqn}
\end{equation}
where, $\rho_m$ and $s$ are the mass density and the specific heat of the material respectively, $T$ is the temperature, $k$ is the thermal conductivity of the material and $Q$ consists of the rate of heat generation per unit volume (as a result of Joule heating) and radiative heat loss. Solution of this equation gives space-time dependence of temperature. The finite element analysis starts with discretizing the film and the substrate into numbers of ``8-node hexahedral" or ``brick elements", the elements for the film sitting exactly over those for the substrate (see FIG~\ref{FilmFEM}). Length and width of the film elements were same with those of the substrate element. Height of each substrate element was same with the thickness of the substrate while that of film varied depending on the position of the element due to roughness of the film surface and they were taken according to the topography of the pristine film. Then following standard FEA procedure~\cite{reddy}, the temperature distribution inside each element was approximated as: 
\begin{equation}
T(x,y,z,t) = \sum_{i=1}^8 T_i^{el}(t)\psi _i^{el}(x,y,z)
\label{basisEqn}
\end{equation}
where, $T_i^{el}(t)$ are the temperatures at the nodes (nodal temperature) of the element at any simulation timestep, $\psi _i^{el}(x,y,z)$ are some appropriate functions, known as shape function in FEA literature, with which temperature anywhere on and inside the element can be written. Using these approximation functions and following `Galerkin weighted residual method'~\cite{reddy}, the equation~\ref{heatFlowEqn} was re-written in a set of eight equations for each element  where nodal temparatures were the unknown variables and the co-efficients were calculated using the shape functions and the thermo-physical parameters used in the equation~\ref{heatFlowEqn}. The dimensions of the elements also enter into these co-efficients at this stage.   Then, assembling all the elements from the film and the substrate, a set of equations was formed for all the nodes, solution of which gives the temperature at each node at any instant. Using the nodal temperature and the equation~\ref{basisEqn}, temperature anywhere inside the film and the substate could be calculated. Temperature at the next simulation time could be calculated using forward time-step finite difference scheme. For this simulation we assumed that both the film and the substrate were thermally isotropic. We used $\rho_m=1.049\times 10^5~kg.m^{-3}$, $s=235.4~Jkg^{-1}K^{-1}$ and $k=430~Wm^{-1}K^{-1}$ for silver film and $\rho_m=2.330\times 10^4~kg.m^{-3}$, $s=700~Jkg^{-1}K^{-1}$ and $k=150~Wm^{-1}K^{-1}$ for silicon substrate. At each simulation step, using the local temperature distribution obtained from FEA and calculating local stress differences, the atomic in-flux and out-flux for each film-element were calculated using the equation~\ref{atomicFluxEq}. This gave the value of change in total number of atom inside the element. Multiplying this number by atomic volume, $\Omega$, change in volume of the element was calculated accordingly change in height also calculated (length and width of each element were kept constant throughout the simulation).  
     
\newpage
%\section{References}

\newpage
\centerline{\large\bf Figure Captions}

\begin{figure}
\vspace{6 cm}
\includegraphics[width=8cm]{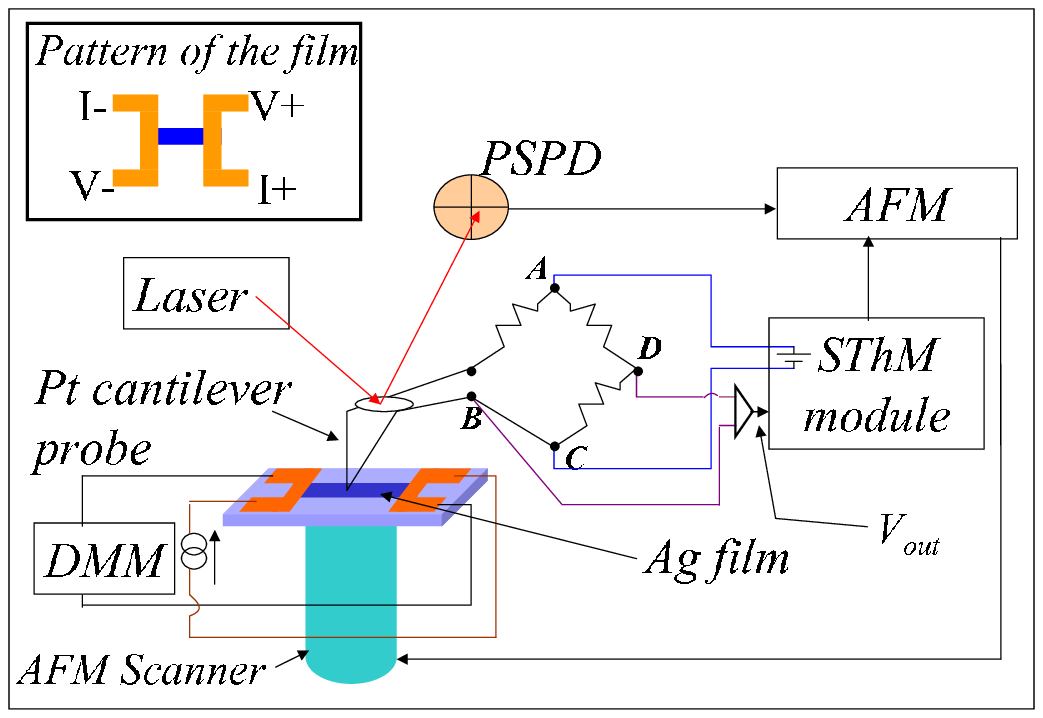}
\caption{}
\label{SThMSchematic}
\vspace{6 cm}
\end{figure}

\begin{figure}
\vspace{6 cm}
\includegraphics{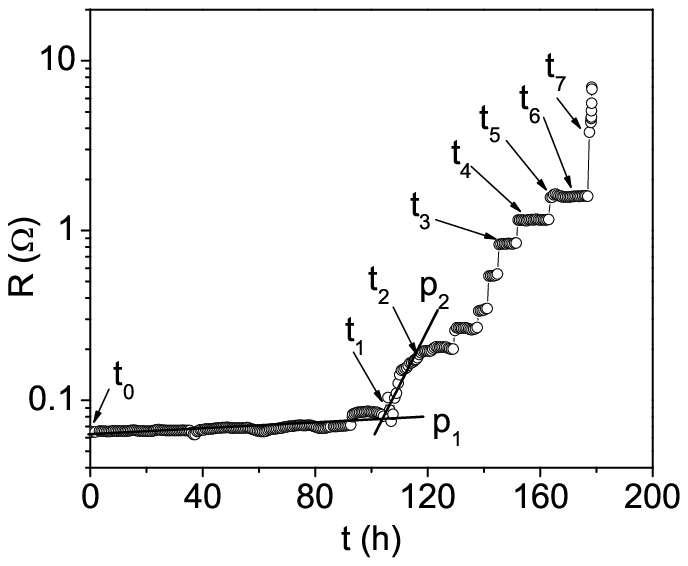}
\caption{}
\label{SThMEMplot}
\vspace{6 cm}
\end{figure}

\begin{figure*}
\vspace{6 cm}
\includegraphics[width=15cm]{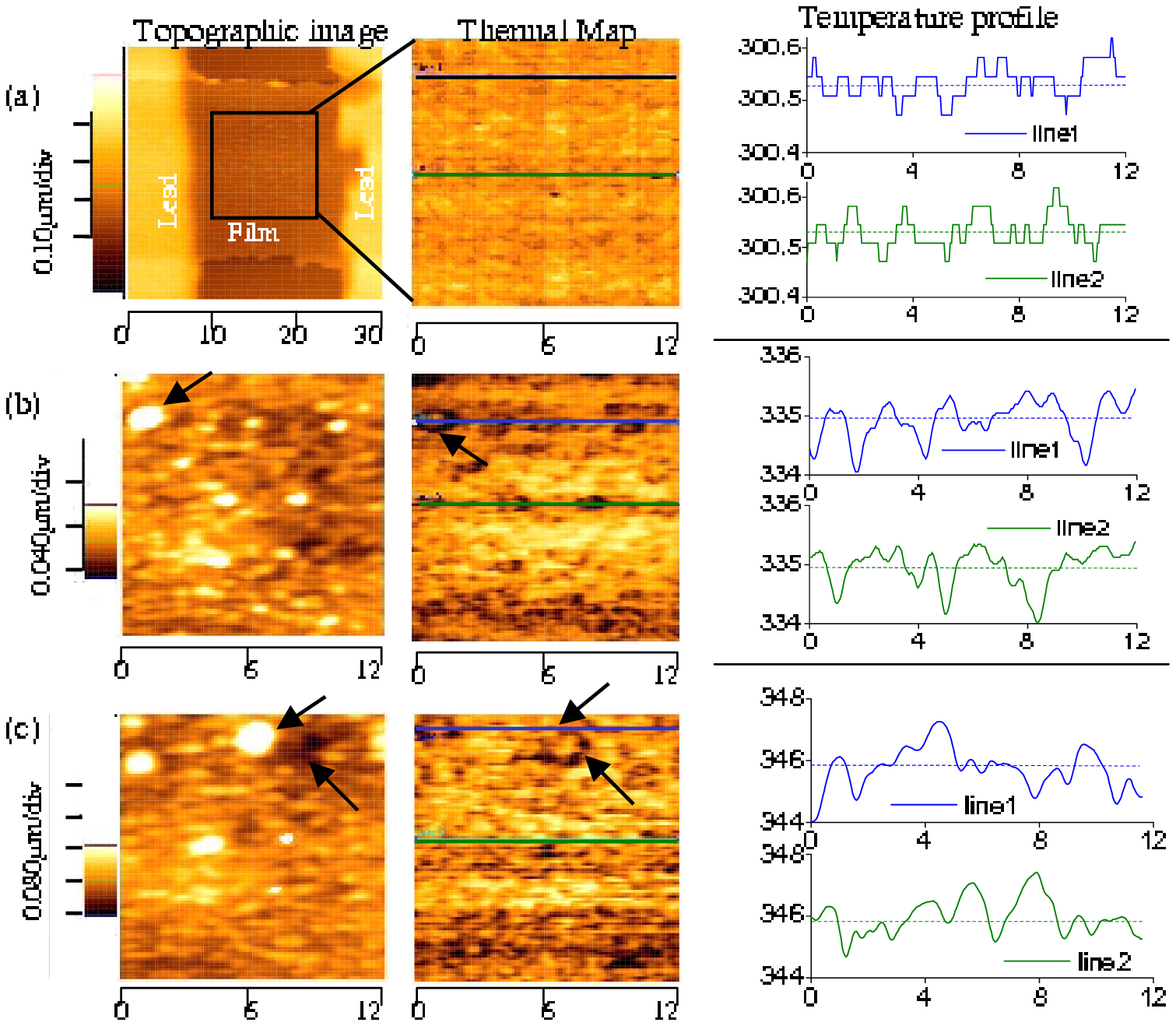}
\caption{}
\label{TopoSThM}
\vspace{6 cm}
\end{figure*}

\begin{figure}
\vspace{6 cm}
\includegraphics{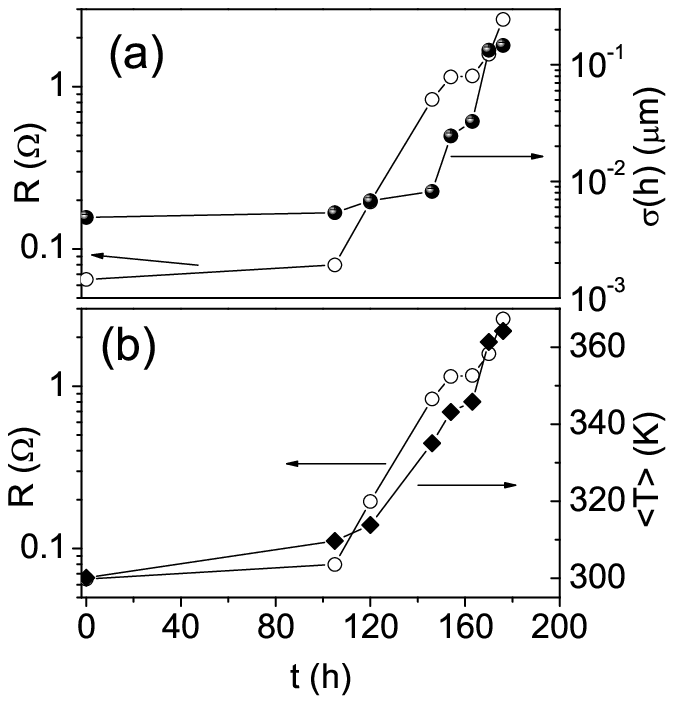}
\caption{}
\label{ResTimeRMSmeanT}
\vspace{6 cm}
\end{figure}

\begin{figure}
\vspace{6 cm}
\includegraphics{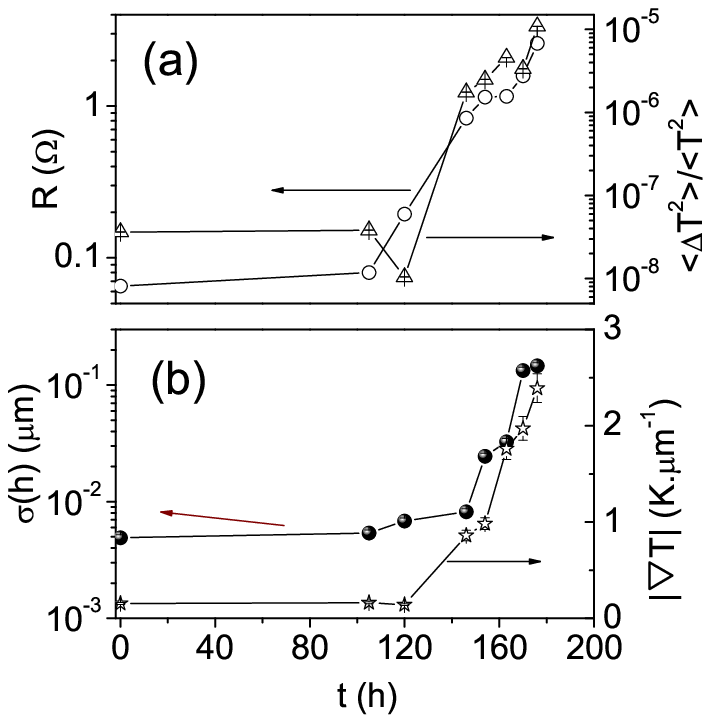}
\caption{}
\label{ResRelVarRMSgrad}
\vspace{6 cm}
\end{figure}

\begin{figure}
\vspace{6 cm}
\includegraphics[width=8cm]{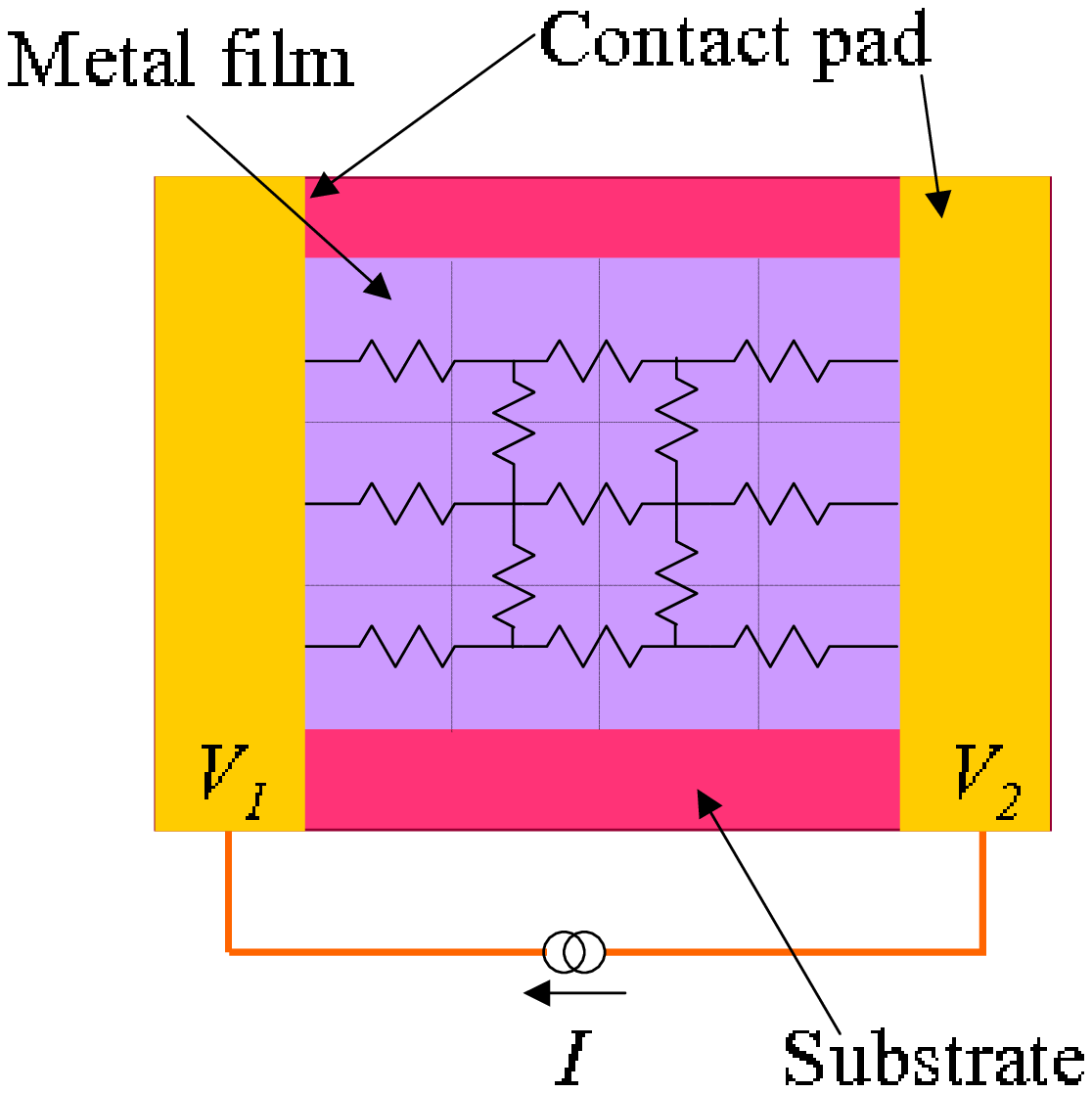}
\caption{}
\label{FilmFEMResNetwork}
\vspace{6 cm}
\end{figure}

\begin{figure}
\vspace{6 cm}
\includegraphics{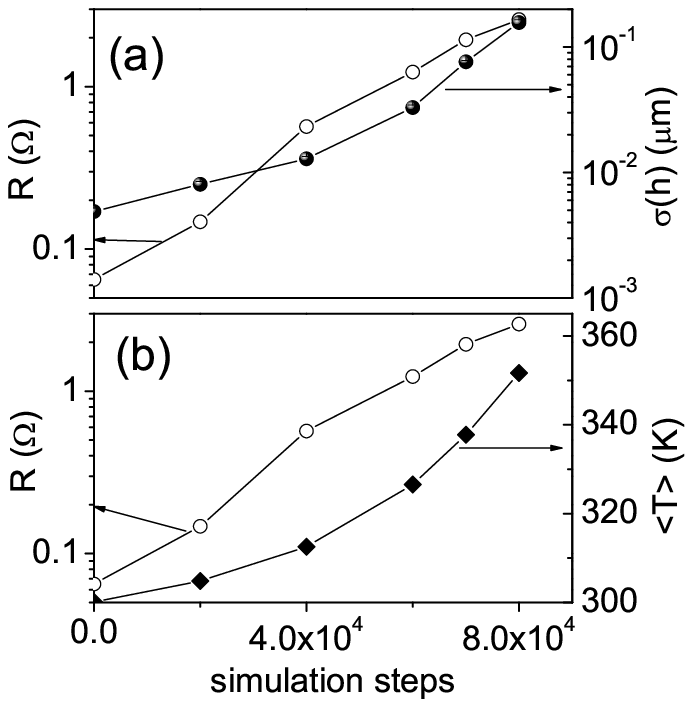}
\caption{}
\label{SimResTimeRMSmeanT}
\vspace{6 cm}
\end{figure}

\begin{figure}
\vspace{6 cm}
\includegraphics{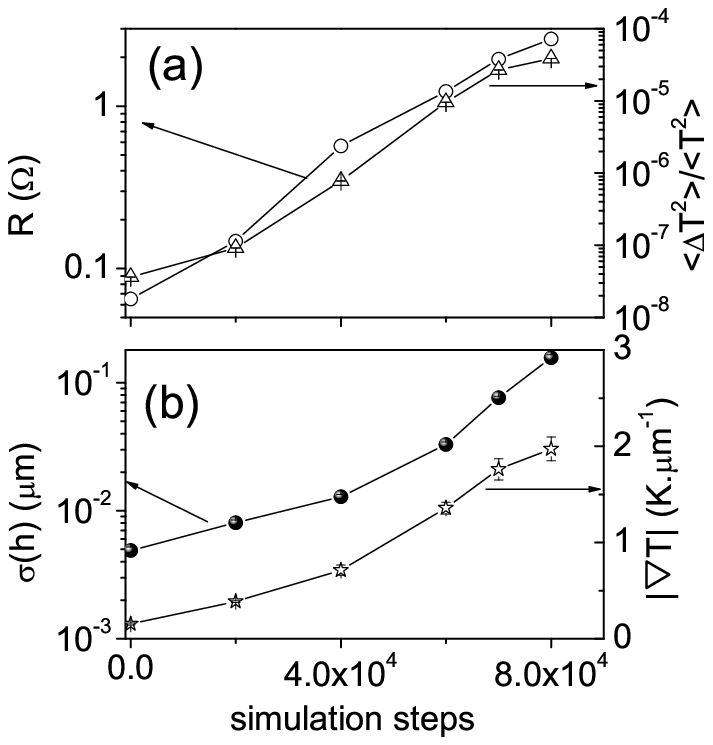}
\caption{}
\label{SimResTimeRelRMSgrad}
\vspace{6 cm}
\end{figure}

\noindent {\bf FIG~\ref{SThMSchematic}:} (Color online) Schematic of experimantal set-up, pattern of the film is shown inset.\\
\noindent {\bf FIG~\ref{SThMEMplot}:} Change in  resistance (R) with time as the damage process progresses. The points marked by $t_0$, $t_1$ etc are the stages for which SThM data are presented here.\\
\noindent {\bf FIG~\ref{TopoSThM}:} (Color online) (Row- wise) Topography, thermal map and typical temperature line profile (a) before current stressing, (b) after 146 hour and (c) after 163 hours for the part of film marked by the rectangle. All lengths are in $\mu m$ and the temperature are in Kelvin scale. Dashed lines are to indicate the mean temperatures along the lines.\\
\noindent {\bf FIG~\ref{ResTimeRMSmeanT}:} Evolution of (a) rms roughness ($\sigma (h)$) and (b) mean temperature ($<T>$) as a function of stressing time. Change in resistance also plotted to show the progress in the damage process.\\
\noindent {\bf FIG~\ref{ResRelVarRMSgrad}:} Change in  (a) resistance (R), relative variances of temperature, (b) rms roughness of the film and local temperature gradient ($\nabla T$) with time as the damage process progresses.\\
\noindent {\bf FIG~\ref{FilmFEMResNetwork}:} (Color online) The film is modeled as a resistance network. For clean presentation only few resistance elements are shown.\\
\noindent {\bf FIG~\ref{FilmFEM}:} (Color online) Modeling of the film and substrate in numbers of finite elements.\\
\noindent {\bf FIG~\ref{SimResTimeRMSmeanT}:} Evolution of (a) rms roughness ($\sigma (h)$) and (b) mean temperature ($<T>$) alongwith resistance as a function of stressing time with the  progress of the damage process as obtained from the simulation.\\
\noindent {\bf FIG~\ref{SimResTimeRelRMSgrad}:} Change in  (a) resistance (R), relative variances of temperature, (b) rms roughness of the film ($\sigma (h)$) and local temperature gradient ($\nabla T$) with time as the damage process progresses as obtained from the simulation.\\

\centerline{\large\bf References}
\begin{references}
\bibitem{TM1} D. R. Campbell and H. B. Huntington,  Phys. Rev. {\bf 179}, 601, (1969).
\bibitem{huntington1} H.B. Huntington, in  Diffusion in Solids: Recent Development, edited by A.S.Nowick and J.J. Burton (Academic Press, New York, 1975), pg. 303.
\bibitem{Blech2} I. A. Blech,  J. Appl. Phys. {\bf 47},1204, (1976).
\bibitem{emrev1} Paul S Ho and Thomas Kwok,  Rep. Prog. Phys. {\bf 52}, 301, (1989).
\bibitem{emrev2} K. N. Tu,  J. Appl. Phys. {\bf 94}, 5451, (2003).
\bibitem{StressTM1} D.Gan, Paul S Ho, R.Huang, J.Leu, J.Maiz  and T. Scherban, J. Appl. Phys. {\bf 97},103531, (2005).
\bibitem{Majumdar} A. Majumdar, K. Luo, J. Lai and Z. Shi, IEEE Prococeedings of International Reliability Physics Symposium, pp 342, (1996).  
\bibitem{Colm} C. Durkan, M.E. Welland, Ultramicroscopy {\bf 82}, 125, (2000).
\bibitem{Joule1} M.L. Trouwborst, S.J. van der Molen and B.J. van Wees, J. Appl. Phys. {\bf 99},114316, (2006).
\bibitem{kondo} S. Kondo and K. Hinode, Appl. Phys. Lett. {\bf 67}, 1606 (1995). 
\bibitem{ju} Y.S. Ju and K.E. Goodson, IEEE Electron Device Lett {\bf 18}, 512 (1997).
\bibitem{igeta}  M. Igeta,  K. Benerjee, G.Wu,  C. Hu and A. Majumdar, IEEE Electron Device Lett. {\bf 21}, 224 (2000).
\bibitem{Veeco} See the manual of CP-Research TM- Microscopes, Veeco Metrology Group, (Veeco Metrology, Inc. 112 Robin Hill Road, http://www.tmmicro.com).
\bibitem{TM2} Y. C. Chuang and C. Y. Liu, Appl. Phys. Lett. {\bf 88}, 174105, (2006).
\bibitem{achyutThesis} Achyut Bora, {\it Ph. D. Thesis}, Department of Physics, Indian Institute of Science, Bangalore, India (2007) 
\bibitem{brune95} H. Brune, K. Bromann, H. R\"{o}der, K. Kern, J. Jacobsen, P. Stoltze, K. Jacobsen and J.K. N\o rskov, {\it Phys. Rev. B}, {\bf 52}, 14380 (1995)
\bibitem{reddy} J.N. Reddy, {\it  Finite element method in heat transfer and fluid dynamics}, (CRC Press, Boca Raton, 2001)
\end{references}
\end{document}